\pdfoutput=1
\documentclass[aps,prl,floatfix,twocolumn,reprint,superscriptaddress]{revtex4-1}
\usepackage{amsfonts}
\usepackage{mathrsfs}
\usepackage{amsmath}
\usepackage{color}
\usepackage{graphicx}
\usepackage{bm}
\usepackage{amssymb}
\usepackage{float}
\usepackage{xspace}
\usepackage{epstopdf}
\usepackage{dcolumn}
\usepackage{longtable}
\usepackage[colorlinks=true, letterpaper=true, pdfstartview=FitV, linkcolor=blue, citecolor=blue, urlcolor=blue]{hyperref}

\begin{document}
\title{From Node-Line Semimetals to Large Gap QSH States in New Family of Pentagonal Group-IVA Chalcogenide }

\author{Run-Wu Zhang}
\affiliation{Beijing Key Laboratory of Nanophotonics and Ultrafine Optoelectronic Systems, School of Physics, Beijing Institute of Technology, Beijing 100081, China}
\affiliation{China Academy of Engineering Physics, Mianyang, Sichuan, 621900, China}

\author{Cheng-Cheng Liu}
\email{ccliu@bit.edu.cn}
\affiliation{Beijing Key Laboratory of Nanophotonics and Ultrafine Optoelectronic Systems, School of Physics, Beijing Institute of Technology, Beijing 100081, China}

\author{Da-Shuai Ma}
\affiliation{Beijing Key Laboratory of Nanophotonics and Ultrafine Optoelectronic Systems, School of Physics, Beijing Institute of Technology, Beijing 100081, China}

\author{Yugui Yao}
\email{ygyao@bit.edu.cn}
\affiliation{Beijing Key Laboratory of Nanophotonics and Ultrafine Optoelectronic Systems, School of Physics, Beijing Institute of Technology, Beijing 100081, China}

\begin{abstract}
Two-dimensional (2D) topological insulators (TIs) have attracted tremendous research interest from both theoretical and experimental fields in recent years. However, it is much less investigated in realizing node line (NL) semimetals in 2D materials.
Combining first-principles calculations and $k \cdot  p$ model, we find that NL phases emerge in \emph{p}-CS$_2$  and \emph{p}-SiS$_2$, as well as other pentagonal IVX$_2$ films, \emph{i.e.} \emph{p}-IVX$_2$ (IV= C, Si, Ge, Sn, Pb; X=S, Se, Te) in the absence of spin-orbital coupling (SOC). The NLs in \emph{p}-IVX$_2$ form symbolic Fermi loops centered around the $\Gamma$ point and are protected by mirror reflection symmetry. As the atomic number is downward shifted, the NL semimetals are driven into 2D TIs with the large bulk gap up to 0.715 eV induced by the remarkable SOC effect.The nontrivial bulk gap can be tunable under external biaxial and uniaxial strain. Moreover, we also propose a quantum well by sandwiching \emph{p}-PbTe$_2$ crystal between two NaI sheets, in which \emph{p}-PbTe$_2$ still keeps its nontrivial topology with a sizable band gap ($\sim$ 0.5 eV). These findings provide a new 2D materials family for future design and fabrication of NL semimetals and TIs.
\end{abstract}
\maketitle
\section{INTRODUCTION}

Semimetals are defined as the systems where conduction bands and valence bands cross each other forming serveral crossing points (CPs) locating near the Fermi level (\emph{E}$_F$) in the Brillouin zone (BZ). Without breaking any of its symmetries, some CPs cannot be removed by perturbations. Among these semimetals, three dimensional (3D) version has drawn broad attentions and achieved a comprehensive undstanding recently~\cite{1,2,3,4,5}. Due to different properties of the CPs, 3D semimetals have been divided into several categories: Dirac semimetals (fourfold degenerate CPs) ~\cite{6,7,8,9,10}, Weyl semimetals (twofold degenerate CPs)~\cite{11,12,13,14,15,16,17}, node line semimetals (the CPs form a closed ring in BZ)~\cite{18,19,20,21,22,23,24},  node chain semimetals (the CPs form a couple of closed and intersected rings) ~\cite{25,26}, node link semimetals (the CPs form a couple of closed and linked rings) ~\cite{27,28,29}, and other node shapes semimetals ~\cite{30,31,32}.

However,  the research field of the two dimensional (2D) counterpart still stays in the developing stage. Lu \emph{et} \emph{al}.~\cite{33} have predicted that 2D mixed lattice composed of honeycomb and kagome lattice (HK lattice) Hg$_3$As$_2$ compound, which presenting 2D NL semimetals due to the band inversion between cation \emph{s} orbital and anion \emph{p}$_z$ orbital. On the other hand, Jin \emph{et} \emph{al}.~\cite{34} also proposed a family group with a zig-zag type monolayer structure of 2D NL semimetals MX (M=Pd, Pt; X=S, Se, Te), and found that these MX compounds host features of 2D NL topological band structures. Furthermore, Yang \emph{et} \emph{al}.~\cite{35} have demonstrated robust NL states in a bipartite square lattice BeH$_2$ and Be$_2$C, which arise from the band inversion by weak SOC. Actually, although structural dynamically and thermally stables of these 2D materials are verfied by phonon calculations and finite-temperature molecule dynamics simulations, it is quite challenging to achieve the freestanding forms in practice. Suitable substrates could not only preserve the intrinsic properties of the 2D materials effectively but provide additional degrees of freedom to design for practical applications. Ideal substrates must satisfy two key factors: (\emph{i}) the substrate is simple and easy-to-implement; (\emph{ii}) the substrate should have a large band gap and should not disturb characteristics of the hosts.

Except for the case of 2D NL enforced by non-symmorphic symmetry~\cite{36}, e.g. a glide mirror, there are some NLs protected by symmorphic symmetry, e.g. a mirror, are vulnerable to the SOC effect. Moreover, SOC effect can drive the symmorphic NLs into 2D TIs. Recently, F. Reis \emph{et} \emph{al}.~\cite{37} have reported that new high-temperature quantum spin Hall (QSH) paradigm in bismuthene on a SiC(0001) substrate in their experimental. A natural question then arises: may SOC effect turn NL phases into QSH phases with a detectable band gap (higher than 26 meV) in some novel 2D materials with high feasibility in reality?

In this work, we use first-principles calculations and $k \cdot  p$ model to put forward a class of 2D NL protected by mirror symmetry (\emph{M}$_z$) in pentagonal IVX$_2$ films, \emph{i}. \emph{e}. \emph{p}-IVX$_2$ (IV= C, Si, Ge, Sn, Pb; X=S, Se, Te) in the absence of SOC. Typical Fermi loops of \emph{p}-IVX$_2$ in \emph{xy} plane show four-fold warping, which is consistent with the \emph{C}$_4$ rotational symmetry and reproduced by our $k \cdot  p$ model perfectly. As the atomic number is downward shifted, the NL semimetals transform into QSH insulators with a sizable bulk gap as large as 0.715 eV induced by SOC, which is identified by \emph{Z}$_2$ topological invariant and dissipationless helical edge states. Noticeably, the inverted band gap in the nontrivial states can be effectively tuned by the biaxial and uniaxial strain. Additionally, we propose a quantum well by sandwiching 2D crystal \emph{p}-PbTe$_2$ between two NaI sheets and reveal that the 2D film remains topologically nontrivial with a sizable gap. These findings may extend our knowledge of 2D NL materials and provide a new platform to design 2D NL and large-gap QSH insulators based on \emph{p}-IVX$_2$ films.

\section{COMPUTATIONAL DETAILS}

To study the structural and electronic properties of \emph{p}-IVX$_2$ films, we employed state-of-the-art ab initio simulations, based on density functional theory (DFT) as implemented in VASP~\cite{38,39}. The exchange correlation interaction was treated within the generalized gradient approximation (GGA) in the form proposed by Perdew-Burke-Ernzerhof (PBE)~\cite{40}, and the plane-wave basis with a kinetic energy cutoff of 300 eV are employed. The Brillouin zone is sampled by using a $9\times9\times1$ Gamma-centered Monkhorst-Pack grid. The vacuum space is set to 20 \AA\ to minimize artificial interactions between neighboring slabs. During the structural optimization of \emph{p}-IVX$_2$ films, all atomic positions and lattice parameters are fully relaxed, and the maximum force allowed on each atom is less than 0.01eV/\AA\ . The screened exchange hybrid density functional by Heyd-Scuseria-Ernzerhof (HSE06)~\cite{41,42} is adopted to further correct the electronic band structure. The phonon calculations are carried out by using the PHONOPY code~\cite{43} through the DFPT approach~\cite{44} without SOC. Here we construct the maximally localized Wannier functions (MLWFs) by employing the WANNIER90 code~\cite{45}. Based on the constructed MLWFs, we obtain the surface Green's function of the semi-infinite system from which we can calculate the dispersion of the edge states.

\section{RESULTS}

\begin{figure}
\includegraphics[width=1.0\columnwidth]{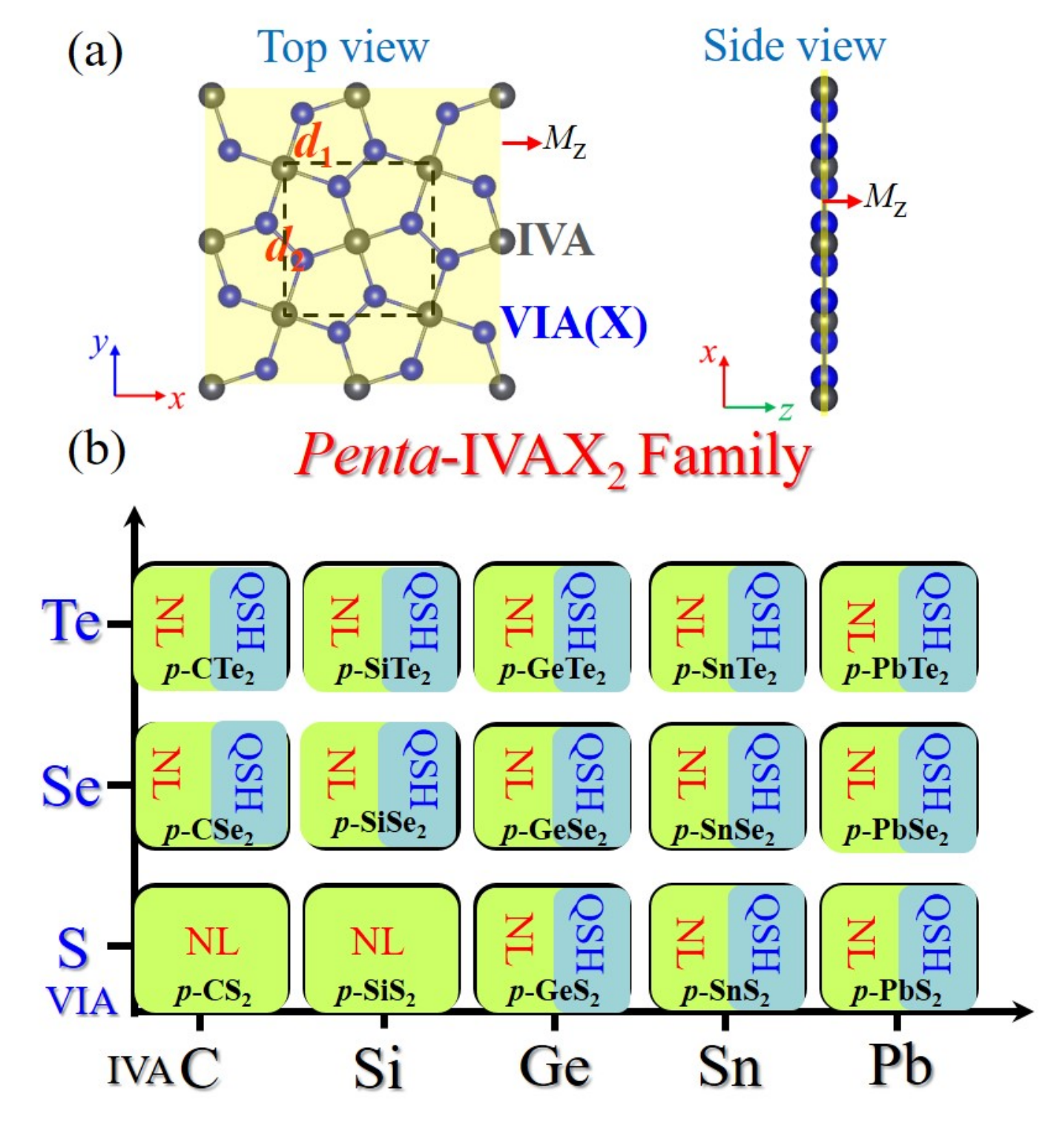}
\caption
{(Color online) (a) Top and side views of the geometrical structures of \emph{p}-IVX$_2$ (IV=C, Si, Ge, Sn, Pb; X = S, Se, Te) films. Black and blue balls denote group IV and group VI atoms, respectively. (b) Summary of the topological states of the whole 15 different \emph{p}-IVX$_2$ based on HSE06 method. In this map, the full-filled light green pattern presents node line (NL) semimetals with negligible SOC. The half-filled light green patterns refer to the NL semimetals in the absence of SOC, while the half-filled light blue notes the QSH insulators in the presence of SOC.}
\label{fig1}
\end{figure}

\emph{p}-IVX$_2$ films possess the square crystal with the space group \emph{P}4/mbm (No.127), which have mirror-inversion planes in \emph{xy} plane (\emph{M}$_z$), that consist of two group IV atoms and four group VI(X) atoms in one unit-cell. The top and side views of \emph{p}-IVX$_2$ as shown in Fig.~\ref{fig1}(a). Moreover, the detailed structural parameters of all the systems are listed in Table \ref{Table1}. It contains the calculated equilibrium lattice constants and bond lengths after the structural optimization. It is known that, from S atom to Te atom (also from C atom to Pb atom), the atomic radius becomes larger in the order of S $<$ Se $<$ Te (C $<$ Si $<$ Ge $<$ Sn $<$ Pb), thus the lattice constants and bond lengths should be increased correspondingly. As shown in the top view, one can see those 2D \emph{p}-IVX$_2$ crystals are akin to the structure of experimentally identified penta-graphene and are composed completely of the pentagonal rings~\cite{46,47,48}. The summary of the topological states of the full 15 different \emph{p}-IVX$_2$ based on HSE06 method is given in Fig.~\ref{fig1}(b). \emph{p}-CS$_2$ and \emph{p}-SiS$_2$ belong to NL with negligible SOC effect. The rest of \emph{p}-IVX$_2$ via SOC undergo the transition from NL semimetals to the TIs.

\begin{table}
  \caption{Calculated structural parameters of the \emph{p}-IVX$_2$ (IV=C, Si, Ge, Sn, Pb; X = S, Se, Te) films, including the lattice parameter \emph{a}({\AA}), the \emph{d}$_1$ and \emph{d}$_2$ are the bond lengths of IV-X and X-X atoms, respectively (in {\AA}), global band gap \emph{E}$_g$ (eV) with SOC effect by PBE and HSE06 methods. M presents metallic of the \emph{p}-IVX$_2$. }
  \label{Table1}
  \begin{tabular}{c|cccc}
    \hline
    \hline
    Structure	& \emph{a}& \emph{d}$_1$ & \emph{d}$_2$ & \emph{E}$_g$(PBE/HSE06)\\
    \hline
    \emph{p}-CS$_2$	    & 5.44	& 2.11	& 2.10	& M	/ M\\
    \emph{p}-CSe$_2$	& 5.99	& 2.29	& 2.50	& M	/ M\\
    \emph{p}-CTe$_2$	& 6.71	& 2.53	&2.98	& M	/ M\\
    \emph{p}-SiS$_2$	& 6.18	&2.45	&2.16	& M	/ M\\
    \emph{p}-SiSe$_2$	& 6.55	&2.54	&2.52	& M	/ M\\
    \emph{p}-SiTe$_2$	& 7.16	&2.76	&2.86	& M	/ M\\
    \emph{p}-GeS$_2$	& 6.42	&2.58	&2.08	& M	/ M\\
    \emph{p}-GeSe$_2$	& 6.77	&2.67	&2.42	& M	/ 0.065\\
    \emph{p}-GeTe$_2$	& 7.36	&2.87	&2.79	& M	/ 0.044\\
    \emph{p}-SnS$_2$	&6.81	&2.77	&2.09	&0.130	/ 0.195\\
    \emph{p}-SnSe$_2$	&7.15	&2.85	&2.43	&0.220	/ 0.345\\
    \emph{p}-SnTe$_2$	&7.74	&3.05	&2.80	&0.119	/ 0.317\\
    \emph{p}-PbS$_2$	&6.99	&2.86	&2.07	&0.406	/ 0.222\\
    \emph{p}-PbSe$_2$	&7.33	&2.94	&2.39	&0.590	/ 0.617\\
    \emph{p}-PbTe$_2$	&7.91	&3.13	&2.77	&0.572	/ 0.715\\
    \hline
    \hline
  \end{tabular}
\end{table}

\begin{figure}
\includegraphics[width=1.0\columnwidth]{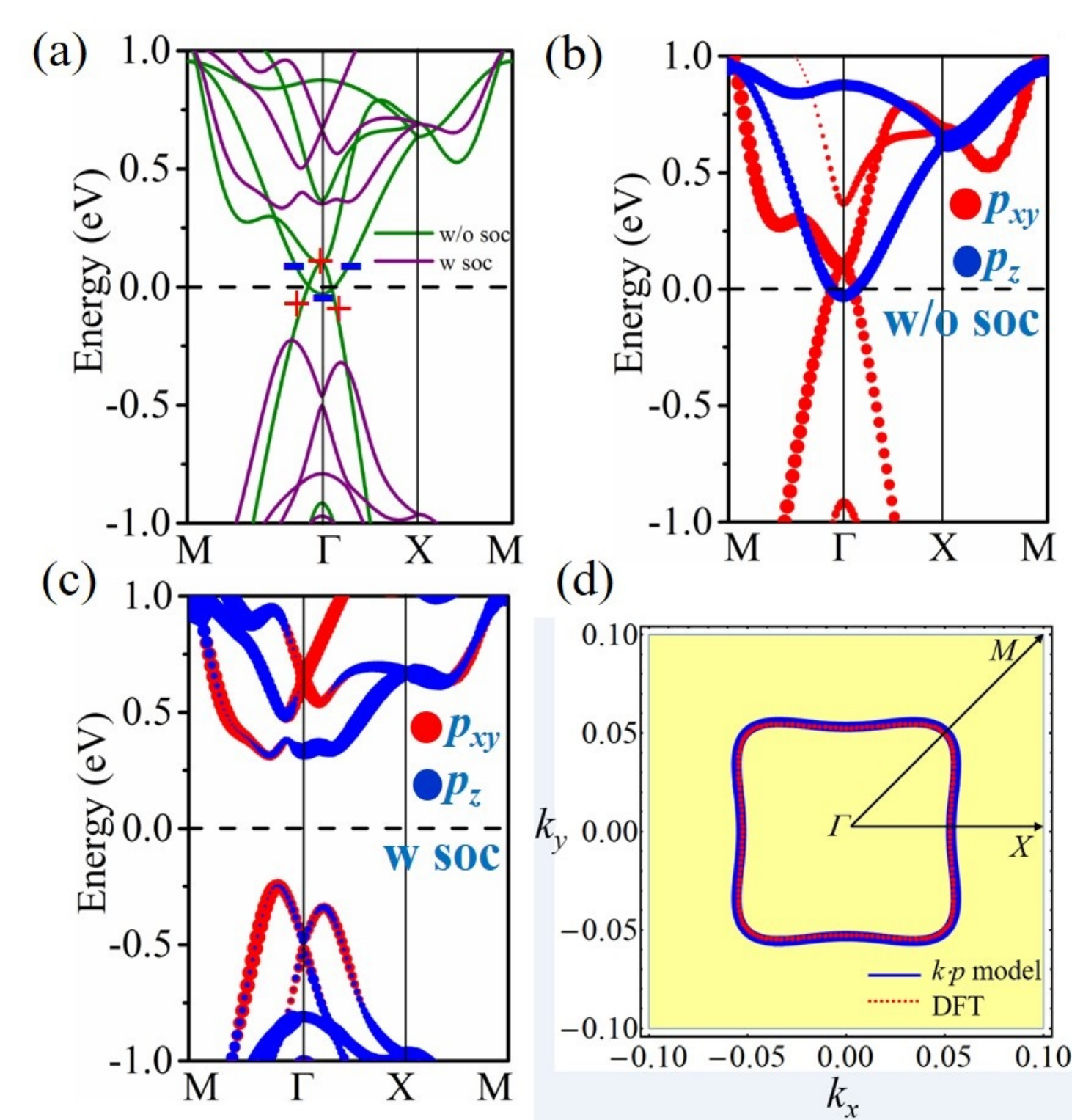}
\caption
{(Color online) The calculated band structures for (a) \emph{p}-PbTe$_2$ film with and without SOC. The olive lines correspond to band structures without SOC, and the purple lines correspond to band structures with SOC. The parity of mirror reflection symmetry for each band is labelled + and - in (a). (b) and (c) Orbital-resolved without and with SOC band structures of \emph{p}-PbTe$_2$, respectively. The blue dots represent the contributions from the \emph{p}$_z$ atomic orbital, and the red dots represent contributions from the \emph{p}$_{x,y}$ atomic orbitals of Pb and Te atoms. (d) A map of consecutive 2D Fermi loops in first Brillouin zone of DFT (red dot) and $k \cdot  p$ model (blue solid) calculations for \emph{p}-PbTe$_2$ without SOC. }
\label{fig2}
\end{figure}

We turn to investigate the electronic properties of freestanding \emph{p}-IVX$_2$. Fig.~\ref{fig2}(a) and supplementary information (Fig.S1) display the band structures for \emph{p}-IVX$_2$, in which the olive and purple lines correspond to band structures without and with SOC. In the absence of SOC, the band structures of \emph{p}-IVX$_2$ show semimetal property along high symmetry directions $M  \to \Gamma \to X$. Interestingly, \emph{p}-IVX$_2$ exhibiting almost linear dispersion of two CPs is viewed located along the high symmetry lines $M  \to \Gamma \to X$ directions, respectively. We find that \emph{p}-PbTe$_2$ and other \emph{p}-IVX$_2$ have similar electronic properties, therefore, we take \emph{p}-PbTe$_2$ as a representative example hereafter. The orbital-projected analysis reveals that valence band maximum (VBM) arise mainly from (Pb \& Te)-\emph{p}$_z$ orbitals, while the conduction band minimum (CBM) is composed of (Pb \& Te)-\emph{p}$_{x,y}$ orbitals, as seen in Fig.~\ref{fig2}(b). At $\Gamma$ point, the system has  time reversal symmetry (\emph{T}), space inversion symmetry (\emph{P}), mirror symmetry (\emph{M}$_z$) and \emph{C}$_4$ rotational symmetry. Given the four symmetries, to the fourth order in \emph{k}, the two-band $k \cdot  p$ Hamiltonians for the NL semimetals may be written as
\begin{equation}
\begin{split}
H(k) = &({a_0} + {a_2}{k^2})\\
&+({d_0} + {d_2}{k^2} + {d_4}{k^4}\cos 4{\theta _k} + {d_4^\prime}{k^4}){\sigma _z},
\end{split}
\end{equation}
where the ${\theta _k} = \arctan \frac{{{k_y}}}{{{k_x}}}$. The $\cos 4{\theta _k}$ term in the Hamiltonian can perfectly capture the four-fold warping of the NL, as shown in Fig.~\ref{fig2}(d). We can see that the Fermi loop calculated by the $k \cdot  p$ model Hamiltonian is compared with the DFT. The $k \cdot  p$ parameters obtained by fitting with the first principle calculations are listed in Table \ref{Table2}. Moreover, we calculate the \emph{M}$_z$ parities located VBM and CBM bands without SOC and find that the parity of VBM band is opposite to that of CBM band, as labeled by the + and -. (see insert of Fig.~\ref{fig2}(a)) The opposing \emph{M}$_z$ parities mean that the gapless NL is protected by mirror reflection symmetry.


In the presence of SOC, the prominent feature is that the band gap is opened with band inversion occurs, \emph{i}.\emph{e}., the components of VBM and CBM states are exchanged clearly, which are opposite to band-state alignment in non-SOC case, as shown in Fig.~\ref{fig2}(c). One can see that the band structure of \emph{p}-PbTe$_2$ produces a semimetal-to-semiconductor transition. Based on PBE method, the sizable global band gap of 0.572 eV is opened. As we known, the PBE method usually underestimates the band gap, the correctional global band gap of \emph{p}-PbTe$_2$ is 0.715 eV via hybrid functional (HSE06). Given the aforementioned symmetries (\emph{T}, \emph{P}, \emph{M}$_z$ and \emph{C}$_4$) and to the second order in \emph{k}, the $k \cdot  p$ Hamiltonians for the 2D TIs near $\Gamma$ point may be written as:
\begin{equation}
\begin{split}
H(k) = &({\tilde a_0} + {\tilde a_2}{k^2}){I_4} + {\tilde b_1}({k_ - }{\sigma _ + } + {k_ + }{\sigma _ - }) \otimes {s_z} \\
&+ ({\tilde d_0} + {\tilde d_2}{k^2}){\sigma _z} \otimes {I_s},
\end{split}
\end{equation}
where ${k_ \pm } = {k_x} \pm i{k_y}$, ${\sigma _ \pm } = {\sigma _x} \pm i{\sigma _y}$, ${I_4}$ and ${I_s}$ are the fourth and second-order unit matrixs, ${s_z}$ and ${\sigma _z}$ are Pauli matrices acting on the real spin and isospin, respectively. The corresponding $k \cdot  p$ parameters obtained by fitting with the first principle calculations are listed in Table \ref{Table2}. The results are also found for the remaining \emph{p}-IVX$_2$ based on HSE06 method, the related values of band gap as illustrated in Table \ref{Table1}, and the corresponding band structures are displayed in supplementary information (Fig.S2). Notably, with the increase of the strength of SOC, we can get abundant NL phases and QSH phases from \emph{p}-IVX$_2$ family. These \emph{p}-IVX$_2$ show different nontrivial band gaps, which is very beneficial for the future experimental preparation and makes it highly adaptable to various application environments.

\begin{table}
  \caption{The parameters for the $k \cdot  p$  Hamiltonian of \emph{p}-PbTe$_2$.}
  \label{Table2}
  \begin{tabular}{c|ccc}
    \hline
    \hline
    w/o SOC	& {\emph{a}$_0$}(eV) &{\emph{a}$_2$}(eV{\AA}$^2$) & {\emph{d}$_0$}(eV)  \\
    	    & 0.0398	& -9.5 & 0.0646	\\
    \hline
            &{\emph{d}$_2$}(eV{\AA}$^2$) &   {\emph{d}$_4$}(eV{\AA}$^4$) & {\emph{d}$_4^\prime$}(eV{\AA}$^4$) \\
    	    & 10.1277 & 3598.5	& 15263 \\
    \hline
    w SOC	& {$\tilde a_0$}(eV) & {$\tilde a_2$}(eV{\AA}$^2$) & {$\tilde b_1$}(eV) \\
 	     & -0.0427	& 13.6	& -1.64	\\
    \hline
     & {$\tilde d_0$}(eV{\AA}$^2$) & {$\tilde d_2$}(eV{\AA}$^2$) &  \\
     & 0.4077 & -22.9& \\
    \hline
    \hline
  \end{tabular}
\end{table}

\begin{figure}
\includegraphics[width=1\columnwidth]{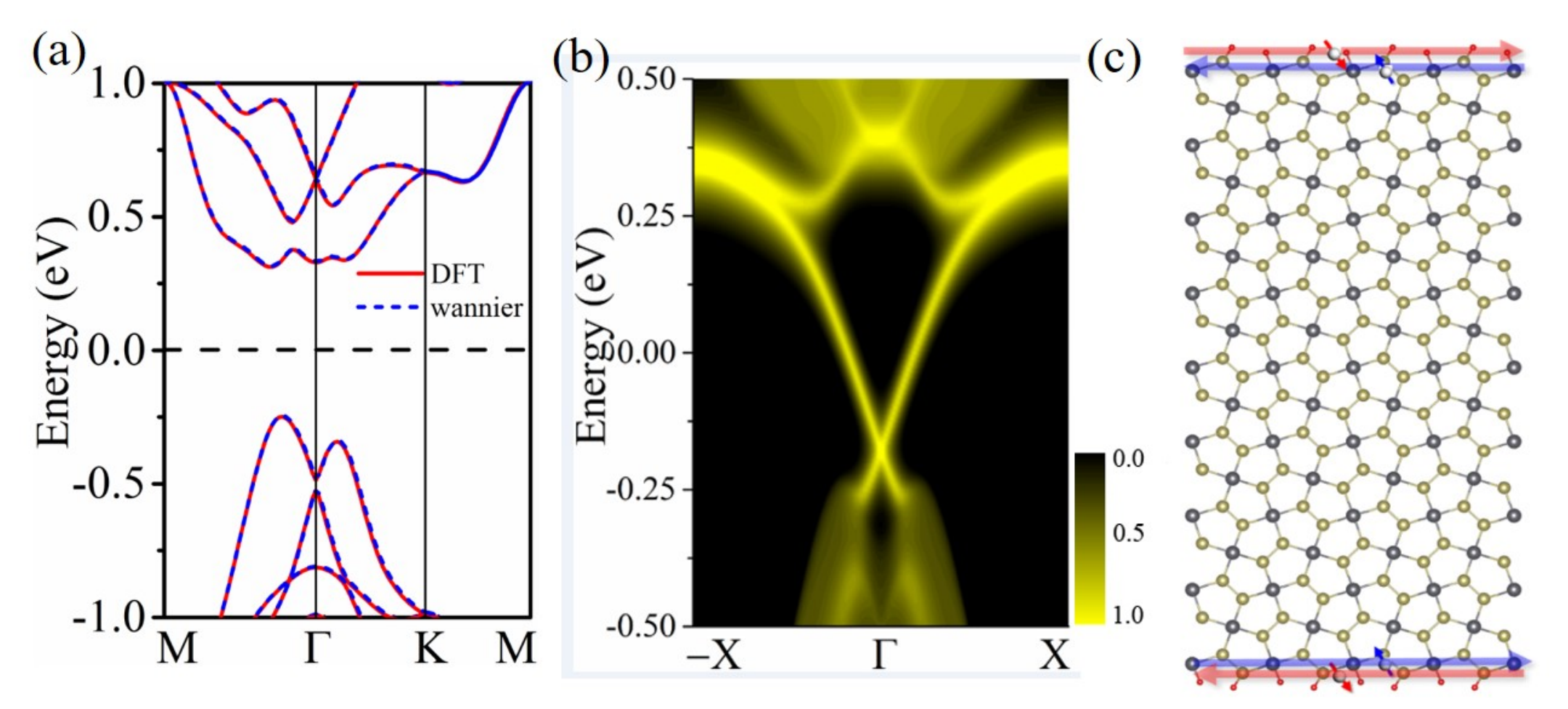}
\caption
{(Color online) (a) A comparison between the band structures of the \emph{p}-PbTe$_2$ calculated from DFT and MLWFs with SOC. (b) Total edge density of states for a semi-infinite \emph{p}-PbTe$_2$. (c) Schematic atomic zigzag-type nanoribbons structure of \emph{p}-PbTe$_2$.}
\label{fig3}
\end{figure}

The helical edge states with spin-polarization protected by time-reversal symmetry (TRS) is the key to identify topological insulators feature. Here, we perform the topological edge states of the \emph{p}-PbTe$_2$ film by the Wannier90 package. Fig.~\ref{fig3}(a) shows the DFT and maximally localized Wannier functions (MLWFs) fitted band structures of the ground state, which are in very good agreement with each other. Then, by using the MLWFs, the spectral function of semi-infinite \emph{p}-PbTe$_2$ is displayed in Fig.~\ref{fig3}(b). We can see that each edge has a single pair of helical edge states in the bulk band gap and cross linearly at the $\Gamma$ point. Obviously, a sizable bulk gap can stabilize the edge states against the interference of the thermally activated carriers, which is valuable for the realizing room-temperature QSH effect. Furthermore, we calculate the \emph{Z}$_2$ invariant $\nu$ based on the method proposed by Fu and Kane~\cite{49}, due to the presence of \emph{p}-PbTe$_2$ film structural inversion symmetry. According to the \emph{Z}$_2$ classification, $\nu$=1 characterizes a topologically nontrivial phase and $\nu$=0 means a topologically trivial phase. Here, the invariants \emph{Z}$_2$ are derived from the parities of wave function at the four TRIM points \emph{K}$_i$, including one $\Gamma$ point and three equivalent \emph{M} points in the Brillouin zone, \emph{i}.\emph{e}., at (0.0, 0.0), (0.5, 0.0), (0.0, 0.5), and (0.5, 0.5) TRIMs. The parity eigenvalue \emph{k} of the $\Gamma$ (0.0, 0.0) points is given by -, while those at the \emph{X} (0.5, 0.0), \emph{Y} (0.0, 0.5), \emph{M} (0.5, 0.5) points remain +, +, +, respectively, yielding a nontrivial topological invariant \emph{Z}$_2$ = 1, which indicates that \emph{p}-PbTe$_2$ films are indeed nontrivial QSH insulator. To demonstrate these features explicitly, we construct a zigzag-type nanoribbon in Fig.~\ref{fig3}(c), whose edge Pb and Te atoms are passivated by hydrogen (H) atoms to eliminate the dangling bonds. The width of this nanoribbon is about 10 nm, which is large enough to avoid interactions between the edge states of the two sides. Notably, the corresponding band structures of \emph{p}-PbX$_2$ nanoribbons are displayed in supplementary information (Fig.S3), and each edge possesses a pair of helical edge states in the bulk band gap and cross linearly at the $\Gamma$ point, which may provide potential applications in realistic nano-electronic devices. Moreover, the sizeable bulk band gaps of \emph{p}-PbX$_2$ nanoribbons can stabilize the edge states against the interference of the thermally activated carriers, which are convenient for observing the nontrivial states in the experiment.

\begin{figure}
\includegraphics[width=\columnwidth]{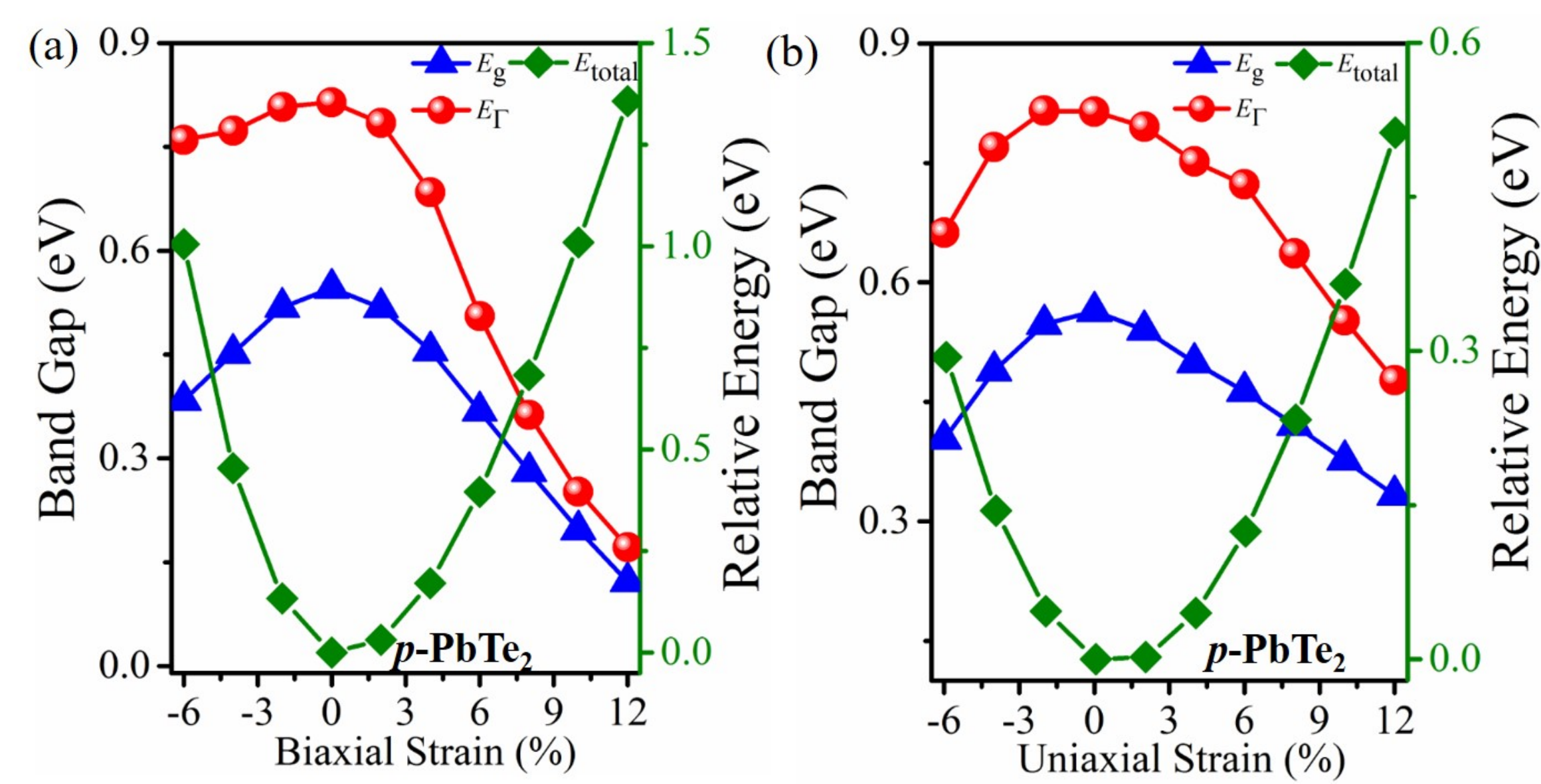}
\caption
{(Color online) (a) and (b) topological phase diagram of \emph{p}-PbTe$_2$ as a function of biaxial and uniaxial strains, respectively. Red/blue dots denote bulk/inverted (local at $\Gamma$ point) band gap. Olive dots show the relative energy of structures under different strains.}
\label{fig4}
\end{figure}

Remarkably, strain engineering provides an effective route to modulate the electronic properties and topological natures, which is meaningful to explore above effects in 2D materials. We apply both external biaxial and uniaxial strain on large gap \emph{p}-PbX$_2$ films by changing its lattices as $\varepsilon  = (a - a{}_0)/{a_0}$, where $a$ ($a{}_0$) is the strained (equilibrium) lattice constants. As shown in Figs.~\ref{fig4}($a-b$), the magnitude of nontrivial band gaps of \emph{p}-PbTe$_2$ can be modified significantly in the range between -6$\% $ to 12$\% $ by biaxial and uniaxial strains, implying the interatomic coupling can modulate the topological natures of these systems. Red and blue dots denote the band gap located at $\Gamma$ point and bulk gap, green dots show the relative energy of structures under different strains. For the biaxial strain of \emph{p}-PbTe$_2$ film, with increasing the strain, it can be seen that both the direct and indirect band gaps in TI phase decrease steadily with respect to compress and tensile strain, leading the band gap of equilibrium state form peak significantly (Fig.~\ref{fig4}(a)). With increasing the lattice constant, the band gap monotonically increases and reaches a maximum value at an equilibrium state, then the band gap decreases. Remarkably, the biaxial strain can produce sizeable nontrivial gaps ranging from 0.122 to 0.572 eV, which is sufficiently large for practical applications at room temperature. While for uniaxial strain (see also Fig.~\ref{fig4}(b) and supplementary information (Fig.S4)), we also can find that QSH states are robust with the uniaxial strain in the wide range of -6 $\sim$ 12 $\% $. Such robust topology against lattice deformation makes \emph{p}-PbX$_2$ easier for experimental realization and characterization on different substrates.

\begin{figure}
\includegraphics[width=1\columnwidth]{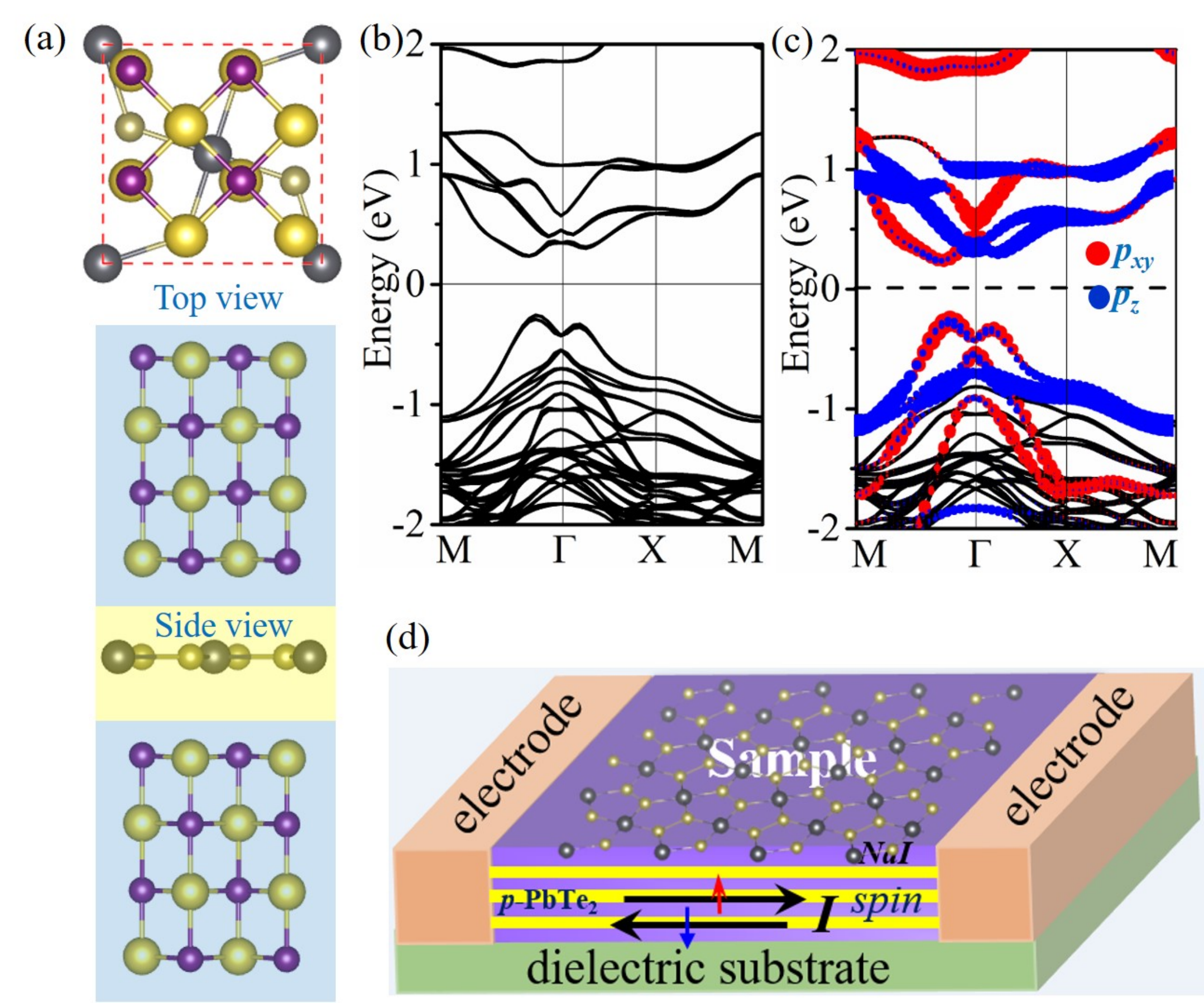}
\caption
{(Color online) (a) Crystal structure of \emph{p}-PbTe$_2$ grown on NaI substrate from the top and side views. (b) band structure and (c) orbital-resolved band structure with SOC for NaI/\emph{p}-PbTe$_2$/NaI heterostructure. (d) Schematic model for proposed NaI/\emph{p}-PbTe$_2$/NaI heterostructure for quantum state measurement. Vertical arrows show the spin orientation of electrons in the edge states and horizontal arrows show their transport directions.}
\label{fig5}
\end{figure}

Starting with the applied perspective, 2D freestanding QSH insulators are quite challenging to achieve in practice. Thus, we take rock-salt NaI with the large gap as an ideal substrate to encapsulate \emph{p}-IVX$_2$ films. Here, we construct the NaI/\emph{p}-PbTe$_2$/NaI heterostructure (HTS) as a typical example. The top and side views of HTS are displayed in Fig.~\ref{fig5}(a). Furthermore, the lattice mismatch of HTS is only about 0.23$\% $, in comparison to NaI (2 $\times$ 2). The optimized distances between adjacent layer are found to be 4.257 {\AA}, and the binding energy is -89 meV, indicating that \emph{p}-PbTe$_2$/NaI is typical van der Waals (vdW) heterostructure. Figs.~\ref{fig5}($b - c$) display the global band structure and projected band structure of \emph{p}$_{x,y}$ and \emph{p}$_z$ orbitals of \emph{p}-PbTe$_2$ on a NaI substrate, respectively. As a result, one can see that robust QSH insulator with the band gap of 0.492 eV, accompanied with band inversion not being affected by the substrate. These results demonstrate the possibility of designing quantum transport devices with this heterostructure, as illustrated in Fig.~\ref{fig5}(d).

\section{CONCLUSION}

In this paper, we predict a class of 2D NL in pentagonal IVX$_2$ films, \emph{i}. \emph{e}. \emph{p}-IVX$_2$ (IV= C, Si, Ge, Sn, Pb; X=S, Se, Te) in the absence of spin-orbital coupling (SOC). The node-line (NL) features are protected by mirror symmetry (\emph{M}$_z$). We can get typical Fermi loops of \emph{p}-IVX$_2$ in \emph{xy} plane from both first-principles calculations and  $k \cdot  p$ model. As the atomic number inceases, the NL semimetals evolve into QSH states with a detectable bulk gap induced by SOC.  Noticeably, the nontrivial band gaps can be effectively tuned by the biaxial and uniaxial strain. Additionally, A vdW heterostructure is bulit to protect the nontrivial topology of the host with a sizable gap. Therefore, \emph{p}-IVX$_2$ and their vdW heterostructures may provide appropriate and flexible candidate materials for realizing low-dissipation quantum electronics and spintronics devices.

\section{acknowledgments}
This work was supported by the National Key R\&D Program of China (No.2016YFA0300600), the MOST Project of China (No.2014CB920903), the National Natural Science Foundation of China (Nos.11774028, 11574029, 11404022), and Basic Research Funds of Beijing Institute of Technology (Grant Nos.20151842003, 2017CX01018).




%

\end{document}